\documentclass[twocolumn,english,aps,prd,graphics,floatfix,a4paper,superscriptaddress]{revtex4-1} 

\usepackage{CJK}
\usepackage{graphicx}
\usepackage{mathrsfs}
\usepackage{bm}
\usepackage{amsmath}
\usepackage{dcolumn}
\usepackage{epstopdf}
\usepackage{dsfont}
\usepackage{amssymb}
\usepackage{tabularx}
\usepackage{array}
\usepackage{float}
\usepackage{color}
\usepackage{epstopdf}
\usepackage{mathrsfs}
\usepackage[colorlinks, linkcolor=blue,anchorcolor=blue,citecolor=blue,urlcolor=blue]{hyperref}

\begin{document}
	
\title{Super-Andreev reflection and longitudinal shift of pseudospin-one fermions}

\author{Xiaolong Feng}
\affiliation{Research Laboratory for Quantum Materials, Singapore University of Technology and Design, Singapore 487372, Singapore}

\author{Ying Liu}
\affiliation{Research Laboratory for Quantum Materials, Singapore University of Technology and Design, Singapore 487372, Singapore}

\author{Zhi-Ming Yu}
\affiliation{Key Lab of Advanced Optoelectronic Quantum Architecture and Measurement (MOE), Beijing Key Lab of Nanophotonics \& Ultrafine
Optoelectronic Systems, and School of Physics, Beijing Institute of Technology, Beijing 100081, China}

\author{Zhongshui Ma}
\affiliation{School of Physics, Peking University, Beijing 100871, China}

\author{L. K. Ang}
\affiliation{Science and Math, Singapore University of Technology and Design, Singapore 487372, Singapore}

\author{Yee Sin Ang}
\affiliation{Science and Math, Singapore University of Technology and Design, Singapore 487372, Singapore}

\author{Shengyuan A. Yang}
\affiliation{Research Laboratory for Quantum Materials, Singapore University of Technology and Design, Singapore 487372, Singapore}
\affiliation{Science and Math, Singapore University of Technology and Design, Singapore 487372, Singapore}

\begin{abstract}
Novel fermions with a pseudospin-1 structure can be realized as emergent quasiparticles in condensed matter systems.
Here, we investigate its unusual properties during the Andreev reflection at a normal-metal/superconductor (NS) interface.
We show that distinct from the previously studied pseudospin-1/2 and two dimensional electron gas models, the pseudospin-1 fermions exhibit a strongly enhanced Andreev reflection probability, and remarkably, can be further tuned to approach perfect Andreev reflection with unit efficiency for all incident angles, exhibiting a previously unknown {super-Andreev reflection effect}. The super-Andreev reflection leads to perfect transparency of the NS interface that strongly promotes charge injection into the superconductor, and directly manifests as a differential conductance peak which can be readily probed in experiment. Additionally, we find that sizable longitudinal shifts exist in the normal and Andreev reflections of pseudospin-1 fermions. Distinct from the pseudospin-1/2 case, the shift is always in the forward direction in the subgap regime, regardless of whether the reflection is of retro- or specular type.
\end{abstract}

\maketitle

\section{\label{intro}Introduction}
	
In condensed matters, due to the strong electron-ion and electron-electron interactions, the electronic spectrum can be dramatically different from that in free space, and can host peculiar emergent fermions at low energy around the Fermi level. A prominent example is graphene~\cite{novoselov2004graphene,castro2009electronicpropertiesofgraphene}, in which the low-energy electrons behave as pseudospin-1/2 fermions as described by the Dirac-Weyl equation in two dimensions (2D). It has been shown that many important properties of graphene can be attributed to this pseudospin-1/2 Dirac fermion characteristics~\cite{castro2009electronicpropertiesofgraphene}. For instance, graphene exhibits the fascinating Klein tunneling effect~\cite{katsnelson2006chiraltunneling}: the pseudospin-1/2 fermion can tunnel through an potential barrier with unit probability at normal incidence, which is a direct consequence of the linear dispersion and pseudospin conservation.

Inspired by graphene, much effort has been devoted to explore novel emergent fermions in materials~\cite{bradlyn2016beyond,bradlyn2017topological,zhang2019catalogue,vergniory2019complete,tang2019comprehensive}. In particular, as the most straightforward generalization, the pseudospin-1 fermion has attracted enormous research interests recently. Such unusual fermion was initially proposed to emerge in several artificial models and cold atom systems~\cite{dora2011latticegenelization,goldman2011topological,paavilainen2016coexisting,shen2010singlediraccone,urban2011barrier,illes2017kleinalphat3}, and was subsequently identified in realistic 2D materials, such as the blue phosphorene oxide~\cite{zhuly2016bpo}, and the monolayers of Mg$_2$C~\cite{wang2018monolayermg2c}, Na$_2$O and K$_2$O~\cite{hua2019tunable}. Previous studies have also unveiled novel quantum transport properties of pseudospin-1 fermions~\cite{fang2016klein,mukherjee2015observation,vicencio2015observation,malcolm2014magneto,luo2019magneto,fang2019nonuniversal,fang2019pseudospin}, such as the {super-Klein tunneling effect}: the pseudospin-1 fermion exhibits perfect transmission through a potential barrier over a large range of incident angle, and, at certain special limit, such range can even extend to all incident angles~\cite{shen2010singlediraccone,urban2011barrier,illes2017kleinalphat3,fang2016klein}.

There exists another intriguing effect when electrons are scattered at a normal-metal/superconductor (NS) interface---the Andreev reflection~\cite{andreev1964}. In Andreev reflection, an incident electron from the normal-metal side is reflected as a hole, and a Cooper pair is transmitted into the superconductor. For gapless semimetal systems such as those with pseudospin-1/2 or pseudospin-1 fermions, the Andreev reflection involves the coupling of the electron-like state in the conduction band to a hole-like state in the valence band, which is akin to the case of Klein tunneling. For the pseudospin-1/2 fermions in graphene, it has been demonstrated that the two processes are closely connected~\cite{beenakker2008colloquium,beenakker2008correspondence}, and they share the same hallmark signature---both Andreev reflection and Klein tunneling occur with unit efficiency at normal incidence. Inspired by the intimate connection between Andreev reflection and Klein tunneling and in view of the super-Klein tunneling effect of pseudospin-1 fermions, a natural question immediately arises:  \emph{Will there be any interesting physics in the Andreev reflection of pseudospin-1 fermions?}

In this work, we answer this question in the affirmative. We study the scattering properties of pseudospin-1 fermions at a NS junction and show that, besides the unit efficiency at normal incidence, pseudospin-1 fermions exhibit perfect transmission for a large range of incident angles. Remarkably, the perfect Andreev reflection, occurs only dominantly at normal incidence in pseudospin-1/2 fermions, can be extended to all incident angles when the Fermi energy and the excitation energy approach certain limits. In analogy with the super-Klein tunneling, this effect may be termed as the \emph{super-Andreev reflection}. We show that the super-Andreev reflection directly leads to distinct signatures in the differential conductance of the NS junction, which can be measured in experiment. Furthermore, we study the longitudinal (Goos-H\"{a}nchen-like) spatial shift which occurs in this interface scattering process. The existence of such longitudinal shift during the Andreev reflection was only discovered recently~\cite{liu2018gooshanchen,yuzhiming2019anomalousspatialshifts}. Here, we show that sizable longitudinal shift in Andreev reflection also exists for pseudospin-1 fermions, and in contrast to pseudospin-1/2 fermions, this shift is always in the forward direction for the subgap regime, regardless of whether the reflection is of retro- or specular type. Our findings reveal another exotic fundamental physical phenomenon of pseudospin-1 fermions, which further enriches the quantum transport physics of 2D electronic systems, and opens up new possibilities of device applications that harness the unusual transport characteristics of novel emergent fermions.

\section{\label{formulism} Model and Approach}
	
\begin{figure}[t]
	\centering
	\includegraphics[angle=0, width=0.5\textwidth]{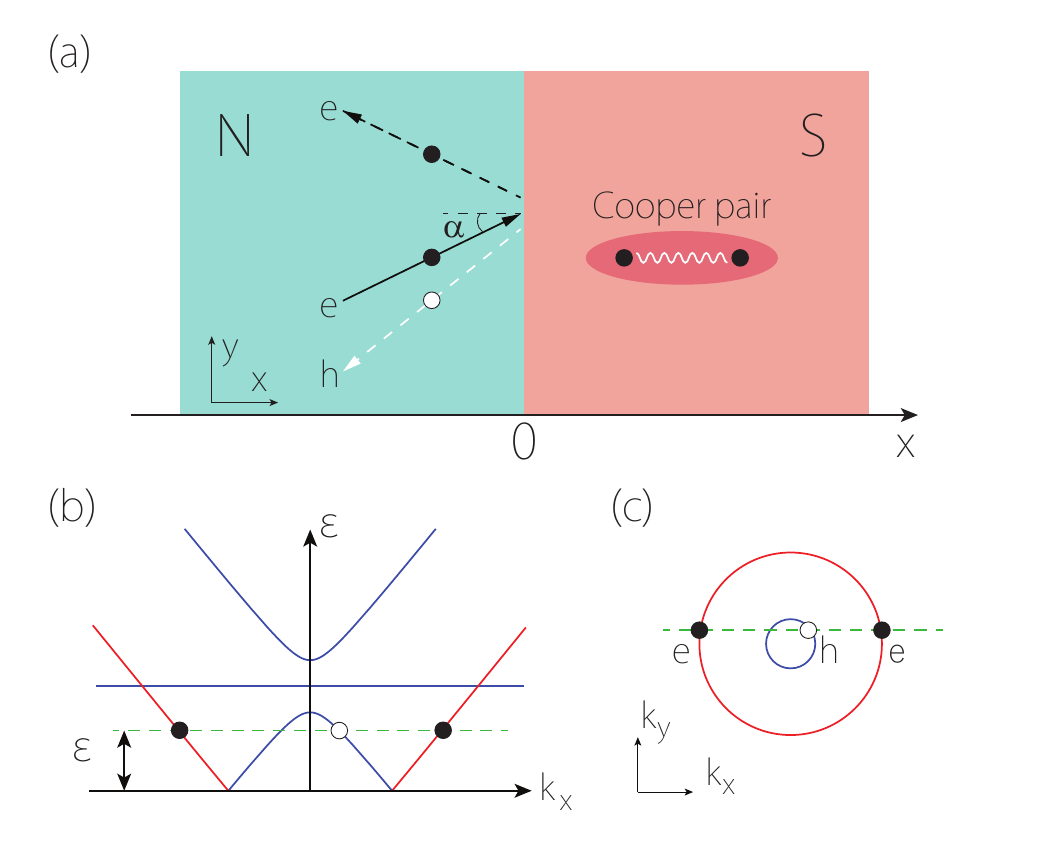}
	\caption{(a) Schematic figure for the NS junction model and the relevant scattering processes. (b) Excitation spectrum in the N region at a constant $k_y>0$. Red (blue) lines denote the electron (hole) bands. The incident electron (the right filled circle) can be reflected as an electron (the left filled circle) or a hole (the open circle). (c) Equienergy contour at the excitation energy $\varepsilon$. }
	\label{AR}
\end{figure}

We consider a general model of a 2D NS junction as shown in Fig.~\ref{AR}(a). The region $x<0$ is the normal metal (N), whereas the region $x>0$ is the superconductor (S). The system is assumed to be uniform and extended in the $y$ direction. The scattering process at the NS interface is described by the Bogoliubov-de Gennes (BdG) equation~\cite{BdG2018},
	\begin{align}
	\begin{bmatrix}
	H_0+U(\bm{r})-E_F&\Delta(\bm{r})\\
	\Delta^*(\bm{r})&E_F-\mathcal{T}^{-1}H_0\mathcal{T}-U(\bm{r})
	\end{bmatrix} \psi = \varepsilon \psi. \label{BdG}
	\end{align}
Here, $\psi$ is the quasiparticle wave function, $\varepsilon$ is the excitation energy with respect to the Fermi level ($E_F$), $H_0$ is the normal-state single-particle Hamiltonian, $\mathcal{T}$ is the time reversal operator, $\Delta(\bm{r})=\Delta_0\Theta(x)$ is the pair potential for the S side, and $\Theta$ the Heaviside step function. In practice, the pair potential can be generated through the proximity effect by covering the 2D material on the S side with a conventional superconductor. The term $U(\mathbf{r})=-U_0\Theta(x)$ represents a potential energy difference between the two sides. In order to satisfy the mean-field requirement of superconductivity, we require $E_F+U_0\gg \Delta_0$, such that the Fermi wavelength in S is much smaller than the coherence length. Meanwhile, the Fermi wavelength on the N side is not constrained to be small, i.e., $E_F$ can be smaller than $\Delta_0$ provided that $U_0$ is large.

For 2D pseudospin-1 fermions, we have the following Hamiltonian (set $\hbar=1$),
\begin{equation}\label{ps1}
  H_0=v_F\bm k\cdot \bm S,
\end{equation}
where $v_F$ is the Fermi velocity, $\bm k=k_x\hat{x}+k_y\hat{y}$ is the wave vector in 2D, and the pseudospin-1 matrices $\bm S$ can be chosen as
\begin{align}
	S_x=\begin{bmatrix}
	0&-i&0\\i&0&0\\0&0&0
	\end{bmatrix},\
	S_y=\begin{bmatrix}
	0&0&i\\0&0&0\\-i&0&0
	\end{bmatrix},\
	S_z=\begin{bmatrix}
	0&0&0\\0&0&i\\0&-i&0
	\end{bmatrix}.
\end{align}
The three $\bm S$ matrices satisfy the angular momentum algebra $[S_i,S_j]=i\epsilon_{ijk}S_k$. The pseudospin-1 fermion is helical, with a helicity of $\pm 1$ and $0$ corresponding to the eigenvalues of the helicity operator $\bm k\cdot \bm S/k$. The two branches with helicity $\pm 1$ form a massless Dirac cone, and the remaining branch with helicity $0$ has a flat dispersion. Note that the pseudospin $\bm S$ has nothing to do with the real electron spin. In the 2D material examples that host such emergent pseudospin-1 fermions~\cite{zhuly2016bpo,wang2018monolayermg2c,hua2019tunable}, the pseudospin comes from the orbital degree of freedom. The real spin in these materials is a dummy degree of freedom, i.e., each branch has an implicit spin degeneracy, due to the negligible spin-orbit coupling. This point will be assumed in our following analysis.

For the purpose of comparison, we also perform the calculation for the pseudospin-1/2 fermions and for the nonrelativistic 2D electron gas model.
The pseudospin-1/2 fermions are described by
\begin{equation}\label{1/2}
  H_0=v_F \bm k\cdot \bm \sigma,
\end{equation}
where $\bm\sigma$ is the vector of the Pauli matrices. This model describes the low-energy electrons in graphene~\cite{castro2009electronicpropertiesofgraphene}. It is worth pointing out that in graphene, there are two valleys located at $K$ and $K'$ points. Nevertheless, via a unitary transformation, the models for both valleys can be put into the same form of Eq.~(\ref{1/2})~\cite{beenakker2006specular}.  The 2D electron gas model is given by
\begin{equation}
  H_0=\frac{k^2}{2m},
\end{equation}
where $m$ is the effective mass. In these two models, the double real spin degeneracy is also implicitly assumed.

The scattering properties of the NS interface are characterized by a set of scattering amplitudes. Consider an electron incident from the N side towards the NS interface. The corresponding scattering state can be written as
\begin{eqnarray}\label{ss}
	\psi(\bm r)=\left\{\begin{split}
	&\psi_{e+}+r_e\psi_{e-}+r_h\psi_{h-},\ \ &x<0&\\
	&t_+\psi_{S+}+t_-\psi_{S-},\ \ &x>0&
	\end{split}\right.
\end{eqnarray}
where $\psi_{e(h)}$ is electron (hole) basis state on the N side, $\psi_{S\pm}$ are the two forward propagating basis states on the S side, the $r$'s and $t$'s are the scattering amplitudes. In this work, we focus on $r_h$, which is the scattering amplitude of the Andreev reflection process.
The scattering amplitudes are solved from the BdG equation (\ref{BdG}) combined with the boundary condition at the interface $x=0$.

\section{\label{model} Super-Andreev reflection}
	
\begin{figure}[t]
		\centering
		\includegraphics[angle=0,width=0.5\textwidth]{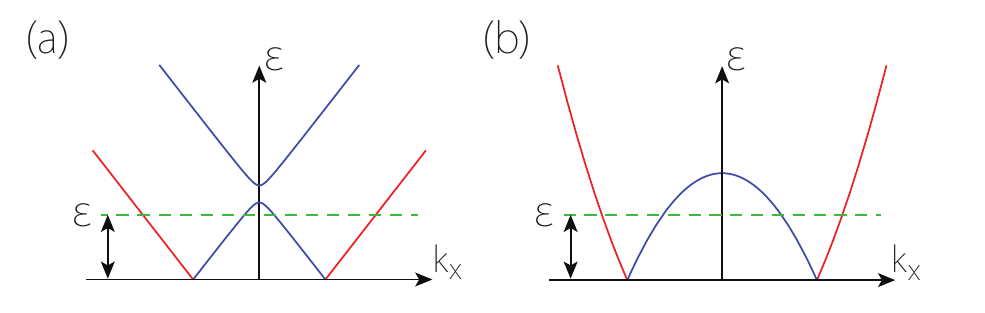}
		\caption{Excitation spectra (in the N region) for (a) the pseudospin-1/2 model, and (b) the 2D electron gas model. Red (blue) lines denote the electron (hole) bands. }
		\label{spec}
\end{figure}

We now proceed to calculate the scattering amplitudes of the NS junction with the pseudospin-1 model in Eq.~(\ref{ps1}). Since the model has particle-hole symmetry, we focus on the case with $E_F\geq 0$. As the junction is assumed to be uniform along the $y$ direction, the wave vector parallel to the interface, i.e., $k_y$, is conserved during scattering. The BdG spectrum for the N side (with a finite $k_y$) is schematically shown in Figs.~\ref{AR}(b) and ~\ref{AR}(c).
Compared with the BdG spectrum of pseudospin-1/2 and 2D electron gas models [see Figs.~\ref{spec}(a) and \ref{spec}(b)], one distinct feature here is the existence of the flatband with helicity $0$, which hosts strongly localized states.
	
The incident electron $\psi_{e+}$ from the dispersive upper band may be reflected to the electron state $\psi_{e-}$ or Andreev reflected to the hole state $\psi_{h-}$. The wave functions for these states can be obtained as
\begin{align}
	&\psi_{e\pm}=\frac{e^{\pm ik^e_x x+ik_yy}}{\sqrt{\cos\alpha}}(1,\pm i\cos\alpha,-i\sin\alpha,0,0,0)^T,\label{psie}\\
	&\psi_{h-}=\frac{e^{-ik^h_x x+ik_yy}}{\sqrt{\cos\alpha'}}(0,0,0,1,i\cos\alpha',i\sin\alpha')^T,\label{psih}
\end{align}
where $\alpha=\arcsin(v_F k_y/\epsilon_+)$ is the angle of incidence, $\alpha'=\arcsin(-v_Fk_y/\epsilon_-)$ is the angle of Andreev reflection, $\epsilon_{\pm}=E_F\pm\varepsilon$, and $k^e_x=\epsilon_+\cos\alpha/v_F$. Note that with finite excitation energy $\varepsilon$, the Andreev reflection only happens when $|\alpha|$ is less than a critical angle
\begin{equation}\label{ac}
  \alpha_c=\arcsin\left| \frac{E_F-\varepsilon}{E_F+\varepsilon}\right|.
\end{equation}
For $|\alpha|<\alpha_c$, $k^h_x=-\epsilon_-\cos\alpha'/v_F$; whereas for $|\alpha|>\alpha_c$, $\psi_{h-}$ becomes an evanescent state with $k^h_x=i[k_y^2-(\epsilon_-/v_F)^2]^{1/2}$, and only the normal reflection is allowed for this case. It is worth noting that $\alpha_c$ approaches $\pi/2$ when $\varepsilon\gg E_F$ or $\varepsilon\ll E_F$. Correspondingly, we have $\alpha'=\alpha$ for $\varepsilon\gg E_F$, and $\alpha'=-\alpha$ for $\varepsilon\ll E_F$. In Eqs.~(\ref{psie}) and (\ref{psih}), the pre-factors $1/\sqrt{\cos\alpha}$ and $1/\sqrt{\cos\alpha'}$ are inserted to ensure the conservation of particle current.

It should be noted that the Andreev reflection here is of retro-reflection type when $\varepsilon<E_F$; and it is of specular reflection type when $\varepsilon>E_F$. This can be easily observed from the sign change of $\alpha'$ as $\epsilon_-$ switches sign. In the retro-reflection case, the reflected hole is in the conduction band; whereas in the specular reflection case, the reflection is in the valence band. As discussed in the case of graphene~\cite{beenakker2006specular}, this specular Andreev reflection is a special feature of semimetal systems.

In the S region, the basis states are given by
\begin{align}\label{psis}
	\psi_{S\pm}=\begin{bmatrix}
	1\\\pm i\cos\gamma\\-i\sin\gamma\\e^{-i\beta}\\\pm i\cos\gamma e^{-i\beta}\\-i\sin\gamma\cos\beta
	\end{bmatrix}e^{\pm ik_0x+ik_yy-\kappa x},
\end{align}
where $\gamma=\arcsin[v_F k_y/(U_0+E_F)]$, $\beta=\arccos(\varepsilon/\Delta_0)$ for $\varepsilon<\Delta_0$ and $\beta=-i\text{ arccosh}(\varepsilon/\Delta_0)$ for $\varepsilon>\Delta_0$, $k_0=[(E_F+U_0)^2/v_F^2-k_y^2]^{1/2}$, and $\kappa=(\varepsilon+U_0)\Delta_0\sin\beta/(v_F^2k_0)$.
	
The boundary condition at the interface $x=0$ is
\begin{align}\label{eqnbd}
	\psi|_{x=0^+}&=\psi|_{x=0^-}.
\end{align}
The scattering amplitudes can be obtained by solving Eqs.~(\ref{ss}-\ref{eqnbd}). The reflection amplitudes for the pseudospin-1 fermions can be analytically solved as
\begin{align} r_e=\frac{\cos\beta(\cos\alpha-\cos\alpha')+i\sin\beta(\cos\alpha\cos\alpha'-1)}{\cos\beta(\cos\alpha+\cos\alpha')+i\sin\beta(\cos\alpha\cos\alpha'+1)},\label{re}\\
r_h=\frac{2\sqrt{\cos\alpha\cos\alpha'}}{\cos\beta(\cos\alpha+\cos\alpha')+i\sin\beta(\cos\alpha\cos\alpha'+1)}.\label{rh}
\end{align}
The probabilities for the two processes are given by $R_e=|r_e|^2$ and $R_h=|r_h|^2$. The explicit expression for the Andreev reflection probability for pseudospin-1 fermions is given by
\begin{equation}\label{R1}
  R_h^\text{ps1}=\frac{4\cos\alpha\cos\alpha'}{(\cos\alpha+\cos\alpha')^2+\sin^2\beta\sin^2\alpha\sin^2\alpha'}.
\end{equation}

\begin{figure}[t]
	\centering
	\includegraphics[angle=0, width=0.5\textwidth]{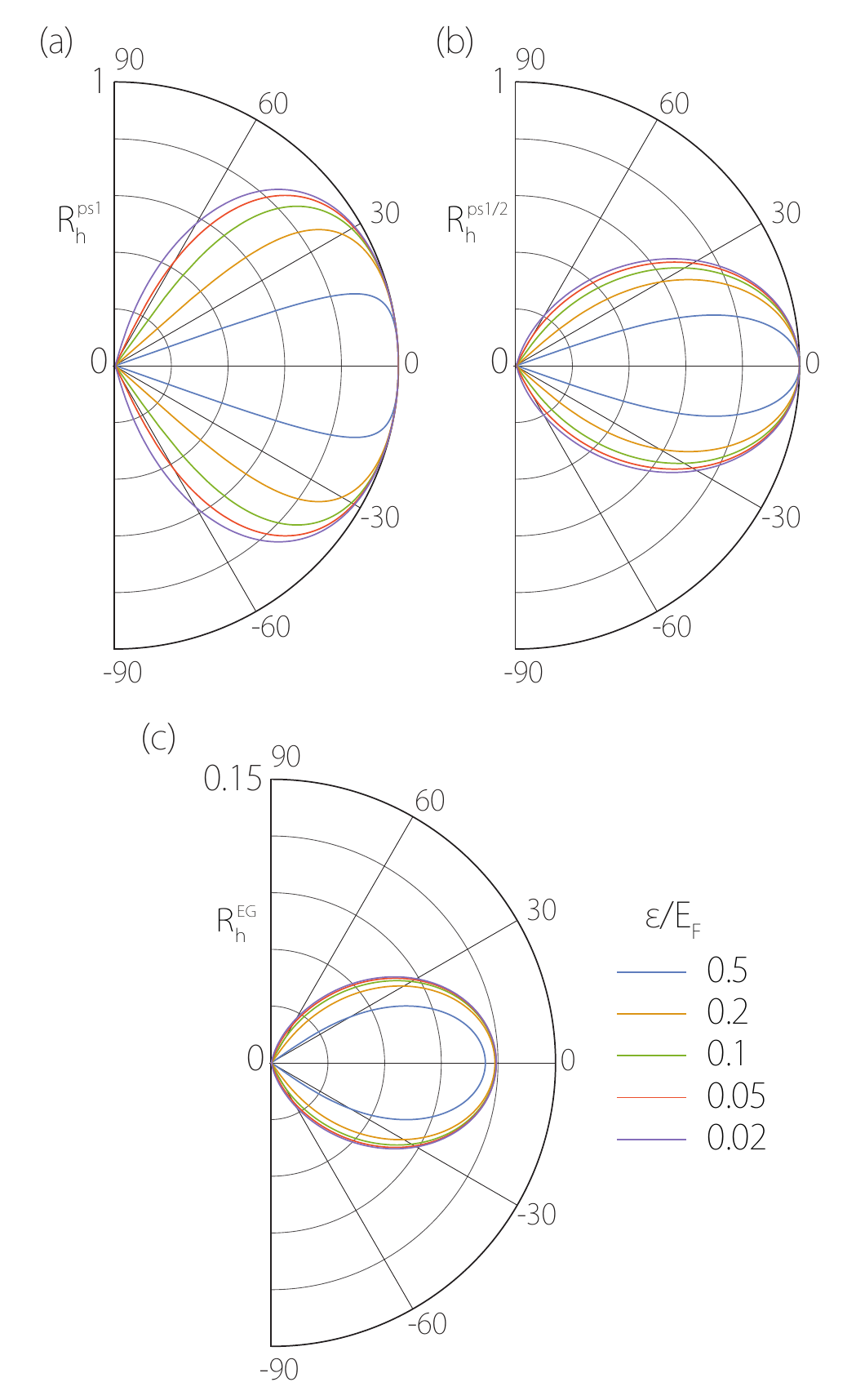}
	\caption{Plots of the Andreev reflection probability $R_h$ as a function of the incident angle $\alpha$, for (a) pseudospin-1, (b) pseudospin-1/2, and (c) 2D electron gas models. The parameters are set as $U_0=1$ eV, $\Delta_0=5$ meV, $v_F=1\times10^6$ m/s, $E_F=3$ meV; For the 2D electron gas model, the effective mass is set to be $m=0.001m_e$.}
	\label{prob}
\end{figure}
	
We observe the following features. First, for the subgap regime with $\varepsilon<\Delta_0$, it is readily verified that $|r_e|^2+|r_h|^2=1$, consistent with the fact that no transmission into S is allowed below the superconducting gap.
Second, at normal incidence, we have $\alpha=\alpha'=0$, such that $R_h^\text{ps1}=1$. This unit efficiency at normal incidence is due to the pseudospin conservation similar to the pseudospin-1/2 case in graphene~\cite{beenakker2006specular}. Third, more importantly, as shown in the plot in Fig.~\ref{prob}(a), nearly perfect Andreev reflection occurs over a large range of the allowed incident angle (note that the incident angle is constrained by $|\alpha|<\alpha_c$).

To further appreciate this enhancement in the Andreev reflection, we compare this result with those for pseudospin-1/2 fermions (as for graphene case) and for 2D electron gas model. Via similar calculations, the Andreev reflection probability for pseudospin-1/2 fermions is given by
\begin{align}
	R_{h}^\text{ps1/2}=&\frac{\cos\alpha\cos\alpha'}{\cos^2\frac{\alpha-\alpha'}{2}-\sin\alpha\sin\alpha'\sin^2\beta},\label{ag}
\end{align}
where $\alpha$, $\alpha'$, and $\beta$ have the same definitions as for the pseudospin-1 case. And for the 2D electron gas model, one finds that
\begin{align}
	R_{h}^\text{EG}=4k_0^2k^e_xk^h_x/\Lambda, \label{adeg}
\end{align}
with
\begin{equation}\begin{split}
  \Lambda=&[k_0(k_x^e+k^h_x)\cos\beta+\kappa(k_x^e-k_x^h)\sin\beta]^2\\
  &+(k_0^2+k_x^e k_x^h+\kappa^2)^2\sin^2\beta.
\end{split}
\end{equation}
Here, except for $\beta$, the symbols are defined with modified expressions compared to the pseudospin-1 case, specifically,
$k_x^e=(2m\epsilon_+-k_y^2)^{1/2}$, $k_x^h=(2m\epsilon_--k_y^2)^{1/2}$, $k_0=[2m(E_F+U_0)-k_y^2]^{1/2}$, and $\kappa=m\Delta_0\sin\beta/k_0$. Note that for this electron gas model, since there is only a single band, the Andreev reflection is only allowed for $\varepsilon<E_F$, and the critical incident angle in this case is given by $\alpha_c=\arcsin\sqrt{\epsilon_- /\epsilon_+}$.

In Figs.~\ref{prob}(b) and ~\ref{prob}(c), we plot the Andreev reflection probabilities for the pseudospin-1/2 fermions and for the 2D electron gas model. For fair comparison, we take $v_F$ to be the same for the pseudospin-1/2 and the pseudospin-1 cases; and for the 2D electron gas model, we take a different Fermi energy from the other two models but require the Fermi velocity at $E_F$ matches $v_F$, i.e., $k_F/m=v_F$.

By comparing the pseudospin-1, pseudospin-1/2 and 2D electron gas in Fig.~\ref{prob}, we arrive at the following results. First, the Andreev reflection probability for the electron gas is generally lower than that of the pseudospin-1 or pseudospin-1/2 fermions. Particularly, at normal incidence, while both pseudospin-1 and pseudospin-1/2 fermions exhibit unit efficiency due to the conservation of pseudospin, the Andreev reflection probability for the 2D electron gas is generally less than one.

Second, the Andreev reflection probability for pseudospin-1 is generally much higher than that of the pseudospin-1/2 fermions. Indeed, by comparing the results in Eq.~(\ref{R1}) and Eq.~(\ref{ag}), one finds that with the same set of parameters, for $\varepsilon<E_F$
\begin{equation}
  R_h^\text{ps1}\geq R_h^\text{ps1/2}
\end{equation}
is always valid.

\begin{figure}[t]
	\centering
	\includegraphics[angle=0, width=0.45\textwidth]{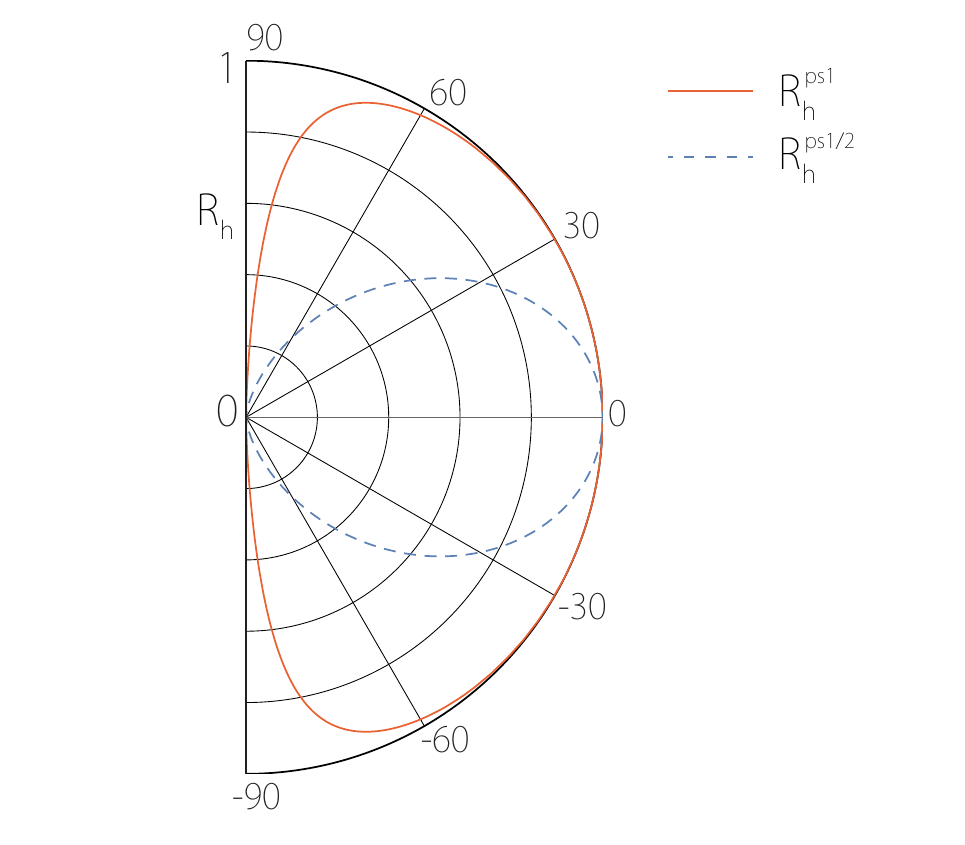}
	\caption{Andreev reflection probability $R_h$ as a function of incident angle $\alpha$ close to the limit of $\varepsilon\rightarrow \Delta_0$ and $E_F\ll \Delta_0$. The red solid (blue dashed) curve is for pseudospin-1 (pseudospin-1/2). The parameters are set as $U_0=1$ eV, $\Delta_0=5$ meV,  $\varepsilon=4.9$ meV, $E_F=0$ meV, and $v_F=1\times10^6$ m/s.}
	\label{plimit}
\end{figure}

Third, for pseudospin-1 fermions, the nearly unit efficiency appears over a wide range of incident angles centered around zero. Such a feature is absent for pseudospin-1/2 fermions in which $R_h$ drops off rapidly away from the zero-angle normal incident case. This behavior can be quantitatively captured by expanding the results in Eq.~(\ref{R1}) and Eq.~(\ref{ag}) at small $\alpha$ around the normal incidence. For pseudospin-1 fermions, we obtain
\begin{align}\label{rh0}
	R_h^\text{ps1}\approx 1-\frac{5+(6+4\sin^2\beta)\lambda^2+5\lambda^4}{16}\alpha^4,
\end{align}
where $\lambda=\epsilon_+/\epsilon_-$. The critical angle $\alpha_c=\arcsin(|\lambda|^{-1})$ ensures that $|\lambda\alpha|<1$ at small incident angles. Eq.~(\ref{rh0}) shows that the deviation from the unit efficiency is small, on the order of $\alpha^4$.
In comparison, for pseudospin-1/2 fermions,
\begin{align}
	R_h^\text{ps1/2}\approx 1-\frac{\lambda^2+(4\sin^2\beta-2)\lambda+1}{4}\alpha^2,
\end{align}
indicating a deviation on the order of $\alpha^2$. Interestingly, we note that the same $\alpha^4$ scaling behavior has also been demonstrated in the super-Klein tunneling effect of pseudospin-1 fermions~\cite{shen2010singlediraccone}.

Furthermore, Eq.~(\ref{R1}) can be simplified in the following limits
\begin{eqnarray}
	R_h^\text{ps1}=\left\{\begin{split}
	&\frac{4\cos\alpha\cos\alpha'}{(1+\cos\alpha\cos\alpha')^2},\ \ \varepsilon\ll \Delta_0,\\
	&\frac{4\cos\alpha\cos\alpha'}{(\cos\alpha+\cos\alpha')^2},\ \ \ \  \varepsilon\rightarrow\Delta_0.
	\end{split}\right.
\end{eqnarray}
Importantly, for $\varepsilon$ close to $\Delta_0$, we have $R_h^\text{ps1}=1$ when $\varepsilon\gg E_F$. In this limit, \emph{perfect} Andreev reflection occurs for all incident angles, giving rise to the \emph{super-Andreev reflection}. In Fig.~\ref{plimit}, we plot the Andreev reflection probability for parameters approaching this limit, indicating the occurrence of super-Andreev reflection. For comparison, we also plot the result for pseudospin-1/2 with the same parameters, which clearly does not exhibit such a behavior.

\begin{figure}[t]
	\centering
	\includegraphics[angle=0, width=0.45\textwidth]{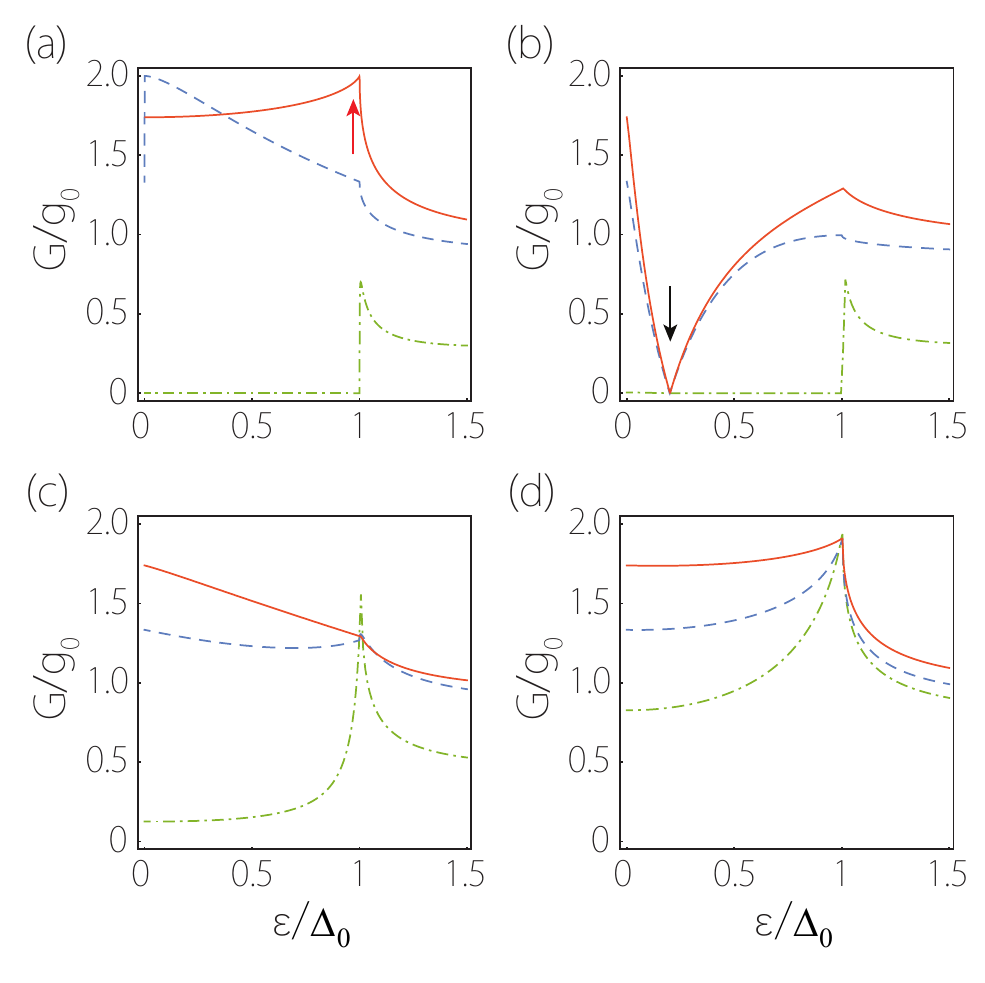}
	\caption{Normalized differential conductances of pseudospin-1 (red solid curve), pseudospin-1/2 (blue dashed curve), and 2D electron gas (green chain dotted curve) NS junctions, with $E_F/\Delta_0$ values given by: (a) 0, (b) 0.2, (d) 5, (e) 50. The red arrow in (a) indicates the signature of the super-Andreev reflection for pseudospin-1 fermions. The black arrow in (b) indicates the point with $\varepsilon=E_F$ where the conductance vanishes for pseudospin-1 and pseudospin-1/2 cases. The parameters are taken as $U_0=1$ eV, $\Delta_0=5$ meV, $v_F=1\times10^6$ m/s, and for the 2D electron gas model, we take $m=0.001m_e$. }
	\label{conductance}
\end{figure}

\section{\label{jcond} Junction conductance}

In an Andreev reflection process, the electric charge of $(-2e)$ is transmitted through the NS interface. Hence, the super-Andreev reflection effect should enhance the charge injection into the superconductor, and is expected to generate strong signatures in the differential conductance measured across the NS junction. The differential conductance of the NS junction can be calculated from the Blonder-Tinkham-Klapwijk (BTK) formula~\cite{blonder1982transition},
\begin{align}
		G=\frac{\partial I}{\partial V}=g_0\int_{0}^{\pi/2}(1-R_e+R_h)\cos\alpha\ d\alpha,
		\label{BTK}
\end{align}
where $g_0$ is the ballistic conductance of the junction in the normal state. By using the reflection probabilities obtained in Sec.~\ref{model}, the differential conductance can be directly evaluated.

In Fig.~\ref{conductance}, we show the differential conductance across the NS junction for a range of $E_F/\Delta_0$ values for the three considered models. One can observe the following features. First, for most cases, the differential conductance for the pseudospin-1 is higher than the other two cases, which reflects an enhanced charge injection into the superconductor and is a directly consequence of the enhanced Andreev reflection of the pseudospin-1 fermions. Second and most importantly, for $\varepsilon\rightarrow\Delta_0$ and small $E_F$ ($E_F\ll\Delta_0$) [see Fig.~\ref{conductance}(a)], $G/g_0$ approaches the maximum conductance value of $2$, indicating that the junction is fully transparent. In contrast, the pseudospin-1/2 and 2D electron gas models do not exhibit such a feature. The perfect transparency of the NS junction in this case thus provides a key experimental signature of the super-Andreev reflection effect. Third, the conductance drops to zero at $\varepsilon=E_F$ for pseudospin-1 and pseudospin-1/2 [see Figs.~\ref{conductance}(b)], because this energy cut through the band degeneracy point where the density of states for propagating states vanishes. Fourth, for large $E_F$ [Fig.~\ref{conductance}(d)], the results for the three cases at $\varepsilon=\Delta_0$ are more or less the same. This is also expected, as for energies away from the pseudospin-1 or pseudospin-1/2 degeneracy point, the electron's behavior should approach that of the free electron model which has a conductance peak at $\varepsilon=\Delta_0$.

In addition, the differential conductance for the junction with pseudospin-1 fermions approaches universal values in the two limits $\varepsilon\ll \Delta_0$ and $\varepsilon\gg\Delta_0$, regardless of the model parameters. This feature is also present for the pseudospin-1/2 fermions in graphene~\cite{beenakker2006specular} and can be readily verified in the calculation. Using the results in Sec.~\ref{model}, we find that for $\varepsilon\ll \Delta_0$, $G\rightarrow(3\sqrt{2}  \text{arctanh} \frac{\sqrt{2}}{2}-2)g_0\approx 1.74 g_0$; whereas for $\varepsilon\gg \Delta_0$, $G\rightarrow(2\pi-16/3)g_0\approx 0.95 g_0$. These values are larger than the corresponding limits for graphene, which are $(4/3)g_0$ for the former limit and $0.86 g_0$ for the latter~\cite{beenakker2006specular}.

\section{\label{shift} Longitudinal shift}

In geometric optics, the longitudinal shift of a light beam within its incident plane during reflection is known as the Goos-H\"{a}nchen effect~\cite{goos1947neuer,de2001evanescent,bliokh2013goos}. Later on, with the development of the field of electron optics, the analogy of this effect in the electronic interface scattering has been discussed in a variety of systems~\cite{sharma2011electron,sinitsyn2005disorder,chen2008tunable,beenakker2009quantum,wu2011valley,chen2013electronic}. More recently, Liu \emph{et al.}~\cite{liu2018gooshanchen} discovered that a longitudinal shift can also exist in the process of Andreev reflection, i.e., the reflected hole beam can have a finite lateral shift with respect to the incident electron beam. In Ref.~\cite{liu2018gooshanchen}, the models for pseudospin-1/2 fermions and for 2D electron gas have been studied in detail. It was found that the pseudospin degree of freedom tends to enhance the longitudinal shift in Andreev reflection.

\begin{figure}[t]
	\centering
	\includegraphics[angle=0, width=0.45\textwidth]{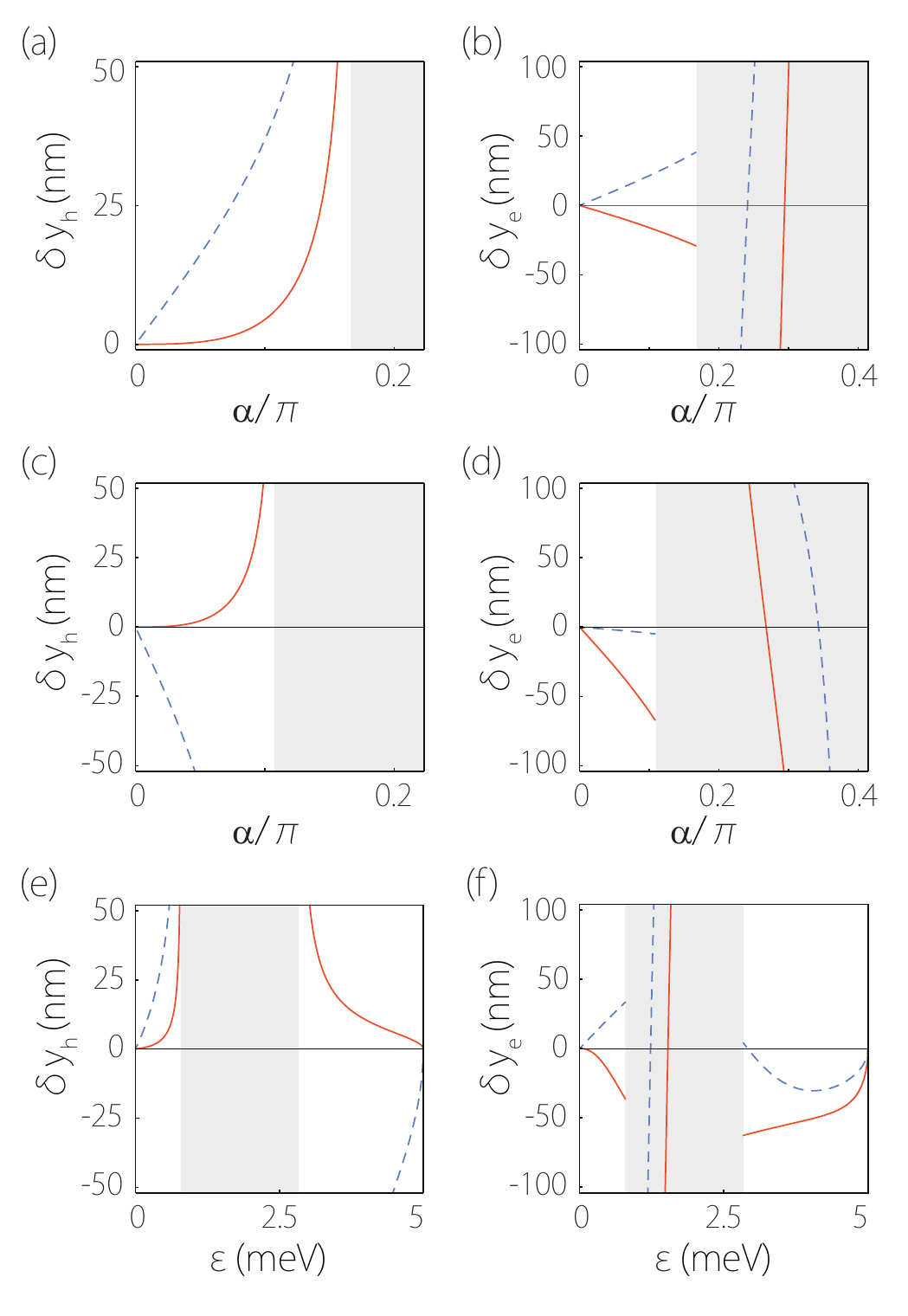}
	\caption{Longitudinal shifts in (a,c,e) Andreev reflection, and  (b,d,f) normal reflection, for pseudospin-1 fermions (red curve) and pseudospin-1/2 fermions (blue curve). The shaded regions correspond to $|\alpha|>\alpha_{c}$, where the Andreev reflection is not allowed.
(a-d) are plotted with respect to the incident angle. (e) and (f) are plotted with respect to the excitation energy.
In the calculations, we set $U_0=1$ eV, $\Delta_0=5$ meV, and $v=1\times10^6$ m/s. For (a,b), we set $E_F=1.5$ meV and $\varepsilon=0.5$ meV. For (c,d), we set $E_F=1.5$ meV and $\varepsilon=3$ meV. For (e,f),  we take $E_F=1.5$ meV, and $\alpha=\pi/10$.}
	\label{ghshift}
\end{figure}

Motivated by these recent findings, we explore the longitudinal shift for pseudospin-1 fermions at the NS interface. As derived in Ref.~\cite{liu2018gooshanchen}, this spatial shift can be directly obtained from the scattering amplitudes. For an incident beam with a central wave vector $(k_x,k_y)$, the shift in Andreev reflection is given by
\begin{align}\label{partialyh}
	\delta y_{h}=-\frac{\partial \text{arg}(r_{h})}{\partial k_y}.
\end{align}
The shift $\delta y_e$ for the normal reflection (i.e., for the reflected electron beam) can be similarly obtained by replacing $r_h$ with $r_e$.
We are most interested in the subgap regime with $\varepsilon<\Delta_0$, where the Andreev reflection can dominate. In this regime, by using the results in Eq.~(\ref{partialyh}), we find that for $|\alpha|<\alpha_c$
\begin{align}
	\delta y_h&= \frac{\sin(2\beta)\sin^2\alpha\sin^2\alpha'\left( \frac{1}{\cos\alpha}+\frac{1}{\cos\alpha'}\right)}{2k_y\big[(\sin\beta\sin\alpha\sin\alpha')^2 + (\cos\alpha+\cos\alpha')^2\big]} \label{hshift}, \\\nonumber
	\delta y_e&= \frac{\sin(2\beta)\sin^2\alpha\sin^2\alpha'\left( \frac{1}{\cos\alpha}-\frac{1}{\cos\alpha'} \right)}{2k_y\big[(\sin\beta\sin\alpha\sin\alpha')^2 + (\cos\alpha-\cos\alpha')^2\big]} \\
	& \ \ \ +\frac{\sin(2\beta)\sin^2\alpha\sin^2\alpha'\left( \frac{1}{\cos\alpha}+\frac{1}{\cos\alpha'}\right)}{2k_y\big[(\sin\beta\sin\alpha\sin\alpha')^2 + (\cos\alpha+\cos\alpha')^2\big]}. \label{eshift}
\end{align}
For $|\alpha|>\alpha_c$, the Andreev reflection is forbidden, while
\begin{align}\nonumber
	\delta y_e=&-\frac{2}{|k^h_x|\sin(2\alpha)\Xi}\Big[2\epsilon_-(\epsilon_+^2\cos^2\alpha+v_F^2|k^h_x|^2\cos2\beta)\\
	&\ \ +v_F|k^h_x|(\epsilon_+^2\sin^2\alpha-2\epsilon_-^2)\sin2\beta\Big],
	\label{acshift}
\end{align}
with
\begin{equation}
  \Xi=\epsilon_+\Big[\epsilon_+^2(1-\sin^2\alpha\sin^2\beta)-v_F|k^h_x|\epsilon_-\sin2\beta-\epsilon_-^2\cos2\beta\Big].
\end{equation}

\begin{figure}[t]
	\centering
	\includegraphics[angle=0, width=0.48\textwidth]{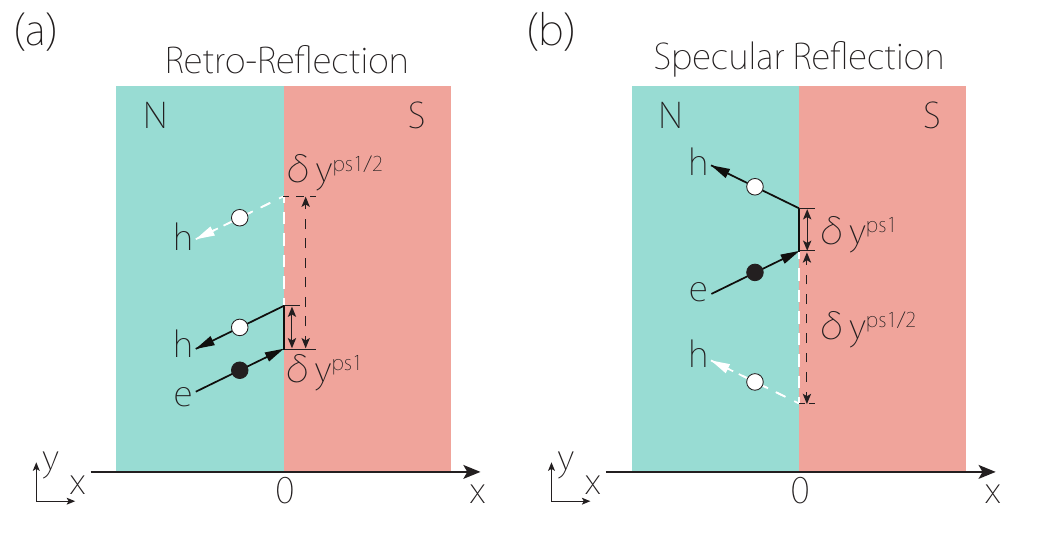}
	\caption{Schematic illustration of the longitudinal shift in Andreev (a) retro-reflection and (b) specular reflection, for pseudospin-1 (black solid line) and pseudospin-1/2 (white dashed line) fermions. For pseudospin-1 fermions, the longitudinal shift is always in the forward direction for the subgap regime. However, for pseudospin-1/2 fermions, the shift is backward for the specular reflection case.}
	\label{degshift}
\end{figure}

These results are plotted in Fig.~\ref{ghshift}. For comparison, we also plot the results for the pseudospin-1/2 case. One observes that sizable longitudinal shifts occur for pseudospin-1 fermions in both Andreev and normal reflections. With the same set of parameters, their values can be comparable to the pseudospin-1/2 case. One distinct feature of pseudospin-1 is that the shift in Andreev reflection is persistently in the forward direction in the subgap regime, regardless of whether the reflection is retro-reflection or specular reflection. This can be easily seen from Eq.~(\ref{hshift}), where the sign of $\delta y_h$ only depends on $k_y$ (all other factors on the right hand side are positive). In stark contrast, for pseudospin-1/2 fermions, its $\delta y_h$ is in the forward direction for retro-reflection ($\varepsilon<E_F$) but in the backward direction for specular reflection ($\varepsilon>E_F$)~\cite{liu2018gooshanchen}. This persistent forward longitudinal shift of pseudospin-1 fermions is schematically illustrated in Fig.~\ref{degshift}.

\section{\label{discussion} Discussion and Conclusion}

In this work, we have revealed several interesting effects in the Andreev reflection of pseudospin-1 fermions at a NS junction. These effects represent nontrivial manifestations of this special pseudospin structure.

We now comment on the experimental aspects. We note that several realistic 2D material candidates, including blue phosphorene oxide~\cite{zhuly2016bpo}, monolayer Mg$_2$C~\cite{wang2018monolayermg2c}, and monolayer Na$_2$O and K$_2$O~\cite{hua2019tunable}, are found to host such pseudospin-1 fermions as low energy excitations. To form a NS interface, the superconductivity may be introduced into these materials by proximity effect from a nearby conventional superconductor. The similar setup has been successfully realized for graphene~\cite{heersche2007bipolar,ojeda2009tuning}. The super-Andreev reflection effect can be detected by measuring the differential conductance of the NS junction, in which a peak conductance of $G/g_0\approx 2$ shall be the signature. The longitudinal shift at the interface can be probed by the schemes proposed in Ref.~\cite{liu2018gooshanchen}. Particularly, the shift would renormalize the group velocity of the waveguide modes in a SNS type waveguide~\cite{liu2018gooshanchen}.

Finally, we mention several related open questions that could be investigated in future works. First, the S side is assumed to have a conventional $s$-wave pair potential in this work. It is interesting to explore the similar effects when pair potential is of unconventional type, e.g., $p$-wave or $d$-wave. The different pairing symmetry should have a strong effect on the Andreev reflection amplitude as well as the anomalous spatial shift~\cite{yu2018unconventionalpairing,yuzhiming2019anomalousspatialshifts}.
Second, besides pseudospin-1, recent works have revealed a variety of novel emergent fermions both in 2D and 3D~\cite{bradlyn2016beyond,bradlyn2017topological,zhang2019catalogue,vergniory2019complete,tang2019comprehensive,yu2019quadratic,wu2019higherorder}. It is certainly of interest to extend the current study to these novel fermions, looking for new physical effects.
	
In conclusion, we have investigated the scattering properties of pseudospin-1 fermions at a NS junction. We find that the Andreev reflection is strongly enhanced. Remarkably, perfect Andreev reflection is allowed over a large range of the incident angle, and at certain limits, perfect Andreev reflection occurs at all incident angles, resulting in the super-Andreev reflection effect. This behavior is in sharp contrast to the pseudospin-1/2 fermions and the 2D electron gas. Experimentally, the super-Andreev reflection strongly promotes charge injection into the superconductor, manifesting as a differential conductance peak, which can be readily probed in experiment. Finally, we show that a sizable longitudinal shift exists in the Andreev reflection. Distinct from the pseudospin-1/2 case, the shift here is persistently in the forward direction, regardless of whether it is retro-reflection or specular reflection.

\begin{acknowledgments}
The authors thank D. L. Deng for valuable discussions. This work is supported by the Singapore Ministry of Education AcRF Tier 2 (MOE2017-T2-2-108). We acknowledge computational support from the National Supercomputing Centre Singapore.
\end{acknowledgments}

\bibliographystyle{apsrev4-1}
\bibliography{spin1}
	
\end{document}